\documentclass[prl,twocolumn,superscriptaddress]{revtex4}

\usepackage{amsfonts}
\usepackage{amsmath}
\usepackage{bbm}
\usepackage{graphicx}
\usepackage{bbold}
\usepackage{amssymb}
\usepackage{color}
\usepackage{subfigure}

\usepackage{mathtools}

\begin{document}

\title{Response to comment on ``Quantum time crystals and interacting gauge theories in atomic Bose-Einstein condensates" by Syrwid, Kosior, and Sacha}

\author{Patrik \"Ohberg}
\affiliation{SUPA, Institute of Photonics and Quantum Sciences, Heriot-Watt University, Edinburgh EH14 4AS, United Kingdom}
\author{Ewan M. Wright}
\affiliation{SUPA, Institute of Photonics and Quantum Sciences, Heriot-Watt University, Edinburgh EH14 4AS, United Kingdom}
\affiliation{Wyant College of Optical Sciences, University of Arizona, Tucson, Arizona 85721, USA}
\begin{abstract}

In a recent comment \cite{SKS} on our paper \cite{OhbWri} Syrwid, Kosior, and Sacha (SKS) asserted that we did not correctly calculate the chiral soliton energy ${\cal E}_{LAB}$ in the lab frame, and further that the system we proposed is not capable of supporting a genuine quantum time crystal.  We concede that we did not correctly calculate ${\cal E}_{LAB}$ but we illustrate below that our system can nonetheless display time crystal behavior with a ground state that is spatially localized and rotating.

\end{abstract}

\maketitle

In the comment by SKS \cite{SKS} the new chiral solitons on a ring that we derived in our paper \cite{OhbWri} are not at issue, but rather whether these can realize a genuine quantum time crystal.  We concede that we used an incorrect expression for ${\cal E}_{LAB}$, but this means that we need to re-examine our results in light of this.  For this purpose we adopt the analysis and notation from the comment by SKS \cite{SKS}, in particular, ${\cal E}_{LAB}$ is given by Eq. (5).

To see how a time crystal can arise first note that an overall phase factor $\exp(i\Theta)$ arises between Eqs. (2) and (3) relating $\Psi(x,t)$ and $\Phi(x-ut,t)$, where
\begin{equation}
\Theta = -{\phi\over 2} + {mux\over\hbar} + {a_1 N\over\hbar}\int^ x ds |\Phi(x-ut,t)|^2 .\nonumber
\end{equation}
Here for our ring geometry $\phi=q\theta$, with $q$ the winding number of the Laguerre-Gaussian laser beam used to induce the gauge potential, and converting to the scaled variables of Ref. \cite{OhbWri} we find $(\sqrt{R}\Phi\rightarrow\varphi)$
\begin{equation}
\Theta(\theta,\tau) = -{q\theta\over 2} + {u\theta\over 2} + a \int^\theta  d\theta'|\varphi(\theta',\tau)|^2. \nonumber
\end{equation}
Then requiring that $[\Theta(\theta+2\pi,\tau)-\Theta(\theta,\tau)]$ is an integer multiple of $2\pi$ so that both $\Psi$ and $\Phi$ are single-valued and obey periodic boundary condition on the ring, and using the normalization condition $\int_0^{2\pi} d\theta' |\varphi(\theta',\tau)|^2=1$, we find for the quantized values of the scaled velocity
\begin{equation}
u = \bigg\{
\begin{array}{cl}
2k - {a\over\pi}, & \mbox{when $q$ is even} \\
(2k-1) - {a\over\pi}, & \mbox{when $q$ is odd}
\end{array}
\nonumber
\end{equation}
where $k=0,\pm 1,\pm 2,\ldots$.  We demand that $\Phi$ obeys periodic boundary conditions in the rotating frame so that the ground state is nodeless and has at most a constant overall phase. The above form for the quantized scaled velocities $u$ extends the discussion in Ref. \cite{OhbWri}.  The key point is that the minimum allowed value of the magnitude of the scaled velocity $|u|$ need not occur at $u=0$.

\begin{figure}
\includegraphics[width=8.5cm]{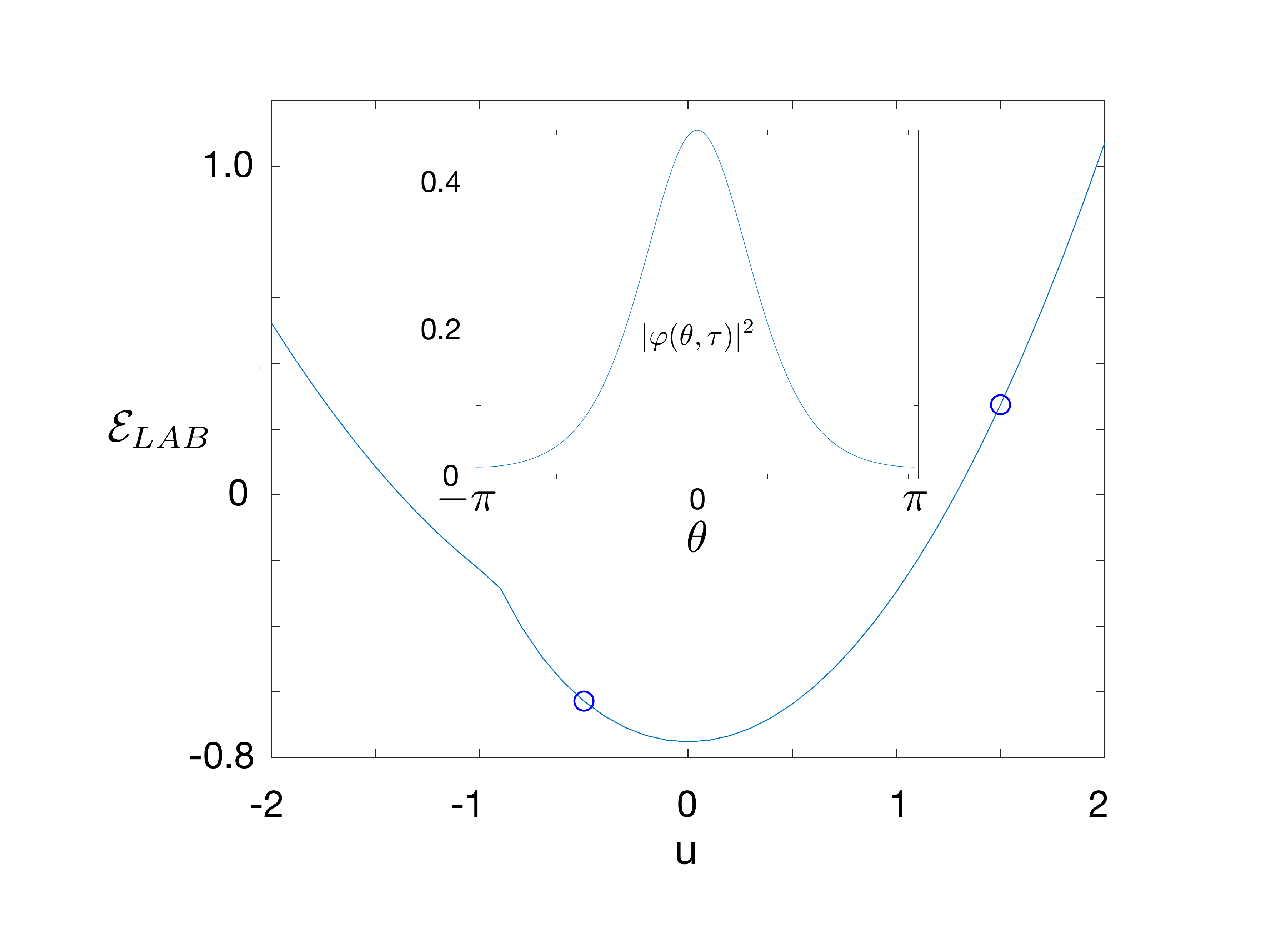}
\caption{Scaled energy ${\cal E}_{LAB}$ in the lab frame versus scaled velocity $u$ for $q$ even, $a={\pi\over 2}$, and $g=-6$. The circles mark the lowest two allowed velocities. The inset shows the spatially localized probability density profile $|\varphi(\theta,\tau)|^2$ in the rotating frame versus $\theta$ for the rotating ground state with $u=-{1\over 2}$.}
\label{figexp}
\end{figure}

As an illustrative example Fig. 1 shows ${\cal E}_{LAB}$ for the chiral soliton versus scaled velocity $u$ for $q$ even, $a={\pi\over 2}$, and $g=-6$:  These calculations employ the scaled version of ${\cal E}_{LAB}$ in Eq. 5 of SKS, and use the exact solutions for $\varphi(\theta,\tau)$ in terms of Jacobi-elliptic functions \cite{Carr2000,KanSaiUed}  that obey the periodic boundary conditions imposed by the ring, as opposed to the hyperbolic-secant approximation used in Eq. 7 of SKS.  The global minimum occurs for $u=0$, but for this example $u=-{1\over 2}$ is the allowed velocity of smallest magnitude, which therefore corresponds to a rotating ground state.  The circles in Fig. 1 mark the lowest two allowed velocities.  But to be a genuine time crystal the probability density profile of the ground state should also be spatially localized.  This can be the case if the scaled parameter $\tilde g=(g-2au)<-\pi$, for which a quantum phase transition occurs leading to spatially localized solutions \cite{OhbWri,KanSaiUed}: The kink in the lab energy curve at $u\sim -0.9$ marks the velocity below which the quantum phase transition to the homogeneous state occurs.  For the above ground state $\tilde g=-4.43$, and the corresponding spatially localized probability density profile $|\varphi(\theta,\tau)|^2$ in the lab frame versus $\theta$ is shown in the inset of Fig. 1.  This illustrative example shows that our system is capable of exhibiting genuine time crystal behavior.  Note that setting $a=0$ in the above example yields $u=0$ for the ground state, so that the current nonlinearity is key in our proposal.  All of these results are in qualitative agreement with Eq. 7 of SKS when account is taken of the quantization of the velocity $u$.

In conclusion, we are grateful to SKS for pointing out that we used the wrong form of the energy of the chiral soliton in the lab frame, but their conclusion that a genuine quantum time crystal can not be formed in our system is not correct.\\

\begin{acknowledgements} 
\noindent P.\"O acknowledges support from EPSRC EP/M024636/1.  
\end{acknowledgements}

\end{document}